\documentclass[twocolumn]{aastex631}

\newcommand{\maya}[1]{{\textcolor{black} {#1}}}

\usepackage{amsfonts}
\usepackage{amsmath,amssymb}
\usepackage[flushleft]{threeparttable}
\usepackage{array,booktabs,makecell}
\begin{document}

\title{The Mass Density of Merging Binary Black Holes Over Cosmic Time}

\author[0009-0003-3908-6112]{Aryanna Schiebelbein-Zwack}
\affiliation{David A. Dunlap Department of Astronomy and Astrophysics, University of Toronto, 50 St George St, Toronto ON M5S 3H4, Canada}
\affiliation{Canadian Institute for Theoretical Astrophysics, 60 St George St, University of Toronto, Toronto, ON M5S 3H8, Canada}

\author[0000-0002-1980-5293]{Maya Fishbach}
\affiliation{David A. Dunlap Department of Astronomy and Astrophysics, University of Toronto, 50 St George St, Toronto ON M5S 3H4, Canada}
\affiliation{Canadian Institute for Theoretical Astrophysics, 60 St George St, University of Toronto, Toronto, ON M5S 3H8, Canada}
\affiliation{Department of Physics, 60 St George St, University of Toronto, Toronto, ON M5S 3H8, Canada}

\begin{abstract}
The connection between the binary black hole (BBH) mergers observed by LIGO-Virgo-KAGRA (LVK) and their stellar progenitors remains uncertain. Specifically, the fraction $\epsilon$ of stellar mass that ends up in BBH mergers and the delay time $\tau$ between star formation and BBH merger carry information about the astrophysical processes that give rise to merging BBHs. We model the BBH merger rate in terms of the cosmic star formation history, coupled with a metallicity-dependent efficiency $\epsilon$ and a distribution of delay times $\tau$, and infer these parameters with data from the Third Gravitational-Wave Transient Catalog (GWTC-3). We find that the stellar progenitors to merging BBHs preferentially form in low metallicity environments with a low metallicity efficiency of $\log_{10}\epsilon_{<Z_t}=-3.99^{+0.68}_{-0.87}$ and a high metallicity efficiency of $\log_{10}\epsilon_{<Z_t}=-4.60^{+0.30}_{-0.34}$ at the 90\% credible level. The data also prefer short delay times. For a power-law distribution $p(\tau)\propto \tau^\alpha$, we find $\tau_\text{min}<1.9 $ Gyr and $\alpha<-1.32$ at 90\% credibility. Our model allows us to extrapolate the mass density in BBHs out to high redshifts, beyond the current LVK detection horizon. We cumulatively integrate our modelled density rate over cosmic time to get the total mass density of merging stellar mass BBHs as a function of redshift. Today, stellar-mass BBH mergers make up only $\sim 0.01\%$ of the total stellar mass density created by high-mass ($>10\,M_\odot$) progenitors. However, because massive stars are so short-lived, there may be more mass in merging BBHs than in living massive stars as early as $\sim 2.5$ Gyr ago. We also compare to the mass in supermassive BHs, finding that the mass densities were comparable $\sim 12.5$ Gyr ago, but the mass density in SMBHs quickly increased to $\sim 75$ times the mass density in merging stellar mass BBHs by $z\sim 1$.
\end{abstract}

\section{Introduction} \label{sec:intro}

Since the first direct detection of gravitational waves (GWs) by the LIGO-VIRGO-KAGRA collaboration (LVK) in 2015, we have entered the catalog era of gravitational wave science \citep{2015CQGra..32g4001L,2015CQGra..32b4001A,2021PTEP.2021eA101A,2016PhRvL.116f1102A,2023ApJ...946...59N,2023arXiv231106061M,2024PhRvD.109b2001A}. The advent of this era allows detected events to be studied holistically, offering deeper insights into the nature of stellar-mass binary black holes (BBHs) as a whole rather than merely focusing on individual detections~\citep{2019PASA...36...10T,2019MNRAS.486.1086M,2022hgwa.bookE..45V}. For instance, population studies have explored the merger rate of BBHs as a function of their masses, spins and merger redshift~\citep[e.g.][]{2016PhRvX...6d1015A,2017ApJ...851L..25F,2017PhRvD..95j3010K,2017PhRvD..96b3012T,2018ApJ...863L..41F,2019ApJ...882L..24A,2020PhDT........14W,2021ApJ...913L...7A,2023arXiv230207289C,2023ApJ...946...16E,2023ApJ...957...37R,2023PhRvX..13a1048A}. 

Current LVK observations can constrain the population of merging BBHs with masses up to $\sim 100\,M_\odot$ and merger redshifts out to $z \sim 1$. The distribution of BBH component masses peaks at $\approx10\,M_\odot$, with another local maximum at $\approx35\,M_\odot$, while the overall BBH merger rate increases with increasing redshift out to the detection horizon at $z\sim1$, similar to the star formation rate (SFR)~\citep{2023PhRvX..13a1048A}. These characteristics can be combined to understand how the mass density contained in merging BBHs evolves as the universe ages.

Stellar-mass BBHs likely originate from massive stars, with some delay time $\tau$ between the formation of the stars and the BBH merger. Hence, we expect the progenitor formation rate to trace the cosmic SFR~\citep[see, e.g.][for a review]{2014ARA&A..52..415M}. Only a small fraction of all stars will become BHs that merge within a Hubble time and this fraction depends on metallicity \citep{2000ARA&A..38..613K,2010ApJ...715L.138B,2011A&A...530A.115B,2012ApJ...749...91F,2024AnP...53600170C}. High metallicities contribute to increased stellar mass loss via line-driven winds, increased radial expansion, and higher natal kicks, which reduce the BBH merger rate and the mass of merging BBHs~\citep[see, e.g.,][for a review]{2024AnP...53600170C}. Consequently, BBH progenitors likely trace low-metallicity star formation, which peaks earlier (at higher redshifts) than the peak of the total SFR at $z\sim2$, although the metallicity-specific SFR is uncertain~\citep{2014ARA&A..52..415M,2019A&ARv..27....3M}.

The distribution of delay times between the formation of the progenitor stars and the merger of the BBH is also uncertain and depends on the astrophysical processes, or formation channels, responsible for merging BBHs ~\citep[see, e.g.][for a review]{2021hgwa.bookE..16M, 2022PhR...955....1M}. Predictions for the delay time distribution are often approximated as power laws $p(\tau)\propto \tau^\alpha$ above some minimum delay time $\tau_\mathrm{min}$. The typical minimum delay time is $\sim10$ Myr, the lifetime of a massive star. A slope of $\alpha=-1$, or a flat-in-log delay time distribution, is a typical prediction from isolated binary evolution when the merger rate is dominated by binaries that undergo common envelope \citep{2012ApJ...759...52D, 2023ApJ...957L..31F}. 
Isolated binary evolution without common envelope (via stable mass transfer or chemically homogeneous evolution) predict longer delay times, hence less negative $\alpha$ \citep{2021ApJ...922..110G, 2022ApJ...931...17V}. In dynamical assembly in dense stellar clusters, BBH mergers that occur inside the cluster can prefer shorter delay times, while binaries that are ejected prior to merger tend to experience longer delay times \citep[e.g.][]{2018PhRvD..98l3005R, 2024arXiv240212444Y}. 

The progenitor formation rate and the delay time distribution set the BBH merger rate as a function of cosmic time. Thus, measurements of the redshift evolution of the BBH merger rate inform the SFR, the cosmic chemical history, and the delay time distribution \citep{2021ApJ...914L..30F, 2023arXiv231017625T,2022ApJ...937L..27M, 2023ApJ...957L..31F,2023arXiv231203316V}. Additionally, by fitting the observed BBH merger rate in terms of these astrophysical parameters, we can infer what the rate should be at higher redshifts, beyond the current LVK detection horizon at $z \sim 1$.

The BBH merger rate is traditionally presented as the time derivative of a number density, with units $\text{Gpc}^{-3} \text{year}^{-1}$~\citep{2023PhRvX..13a1048A}. Integrating this merger rate from the beginning of the Universe through today gives the present-day number density of BH merger products~\citep{2021ApJ...914L..18D}. In this work, we combine the BBH merger rate with the measured mass distribution in order to derive the mass density that is contained in BBHs that have merged. This BBH mass rate density is analogous to the SFR, which is conventionally given in units of $M_\odot \text{Gpc}^{-3} \text{year}^{-1}$. Our framework provides an intuitive understanding of how much mass is contained in, and being added to, the BBH mass budget as a function of cosmic time. Comparing the BBH mass budget to the mass contained in stars and other forms of matter is interesting in the cosmological context~\citep{1998ApJ...503..518F}. For example, we can directly compare the mass density in BBH mergers to that of massive stars, seeing whether the SFR can account for the BBH merger rate or if more exotic formation scenarios, like primordial BHs, are favoured~\citep[e.g.][]{2016PhRvL.116t1301B,2022ApJ...933L..41N,2023PhRvD.107j1302M,2024PhR..1054....1C}. Furthermore, we can begin to test the hypothesis that mergers of stellar mass BBHs serve as the seeds to supermassive BHs (SMBHs) by comparing the mass contained in stellar-mass versus supermassive BHs at high redshifts \citep{2020ARA&A..58...27I}. 
\maya{Our approach is complementary to recent work by~\citet{2022ApJ...924...56S,2022ApJ...934...66S}, who predicted the BH mass density from the stellar-mass to supermassive regime from stellar and binary evolutionary models. Rather than forward modeling the merging BBH population, we take the LVK observations as our starting point.}

The remainder of this paper is structured as follows. In Section \ref{sec:LIGO} we compute the mass density rate of LVK BBH mergers using data from the Third Gravitational-Wave Transient Catalog (GWTC-3) \citep{2023PhRvX..13d1039A}. We extrapolate this quantity to higher redshifts $z > 1$ in Section \ref{sec:model} with a physically informed model that we fit to the low-redshift BBH observations. By doing so, we infer properties such as the metallicity-dependent efficiency and time delay distribution. We compare our cumulatively integrated BBH merger mass density to that of their progenitor population, massive stars, in Section \ref{sec:comparison}. Similarly, we compare to the mass density in SMBHs and determine what percentage of SMBH mass could be attributed to stellar mass BBH mergers.   

\section{How much mass is there in LVK's BBH Mergers?}\label{sec:LIGO}
The SFR is conventionally given in units of $M_\odot \text{Gpc}^{-3} \text{year}^{-1}$. We compute a similar quantity for LVK's BBHs:  the mass of merging BBHs that is being added in a comoving volume over a span of time. This differs from the standard BBH rate densities in the literature which are reported in terms of number densities with units $\text{Gpc}^{-3} \text{year}^{-1}$~\citep{2023PhRvX..13a1048A}. 

We begin by fitting the differential merger rate density as a function of BBH mass, spin and redshift. We use the 69 confident BBH events with false alarm rate FAR $< 1$ year and both masses above $3\,M_\odot$ from GWTC-3. We do not consider the non-detection of the stochastic GW background as previous work has shown it has a negligible effect compared to resolved coalescences \citep{2023arXiv231017625T}.  We assume masses, spins and merger redshifts follow independent distributions, so that the merger rate density (number of mergers per comoving volume per source-frame time) follows:
\begin{equation}
\frac{dR}{dm_1 dm_2 d\chi_\mathrm{eff}}(z, m_1, m_2, \chi_\mathrm{eff}) = R(z) p(m_1, m_2)p(\chi_\mathrm{eff}),
\end{equation}
where $R$ is the number of mergers per comoving volume and source-frame time, $p(...)$ represents a normalized probability distribution, $m_1$ is the primary (heavier) component BH mass, $m_2$ is the secondary mass, $\chi_\mathrm{eff}$ is the effective inspiral BBH spin, and $z$ is the merger redshift. In particular, we assume that the BBH mass distribution does not evolve with redshift, which is consistent with GWTC-3 observations~\citep{2021ApJ...912...98F,2023MNRAS.523.4539K}.

We apply the same parametric models from~\citet{2023ApJ...957L..31F} to describe the BBH mass distribution $p(m_1, m_2)$ and spin distribution $p(\chi_\mathrm{eff})$.
The mass distribution follows a slightly modified version of the \textsc{Power Law + Peak} model with smoothing applied at the low- and high-mass ends of the primary mass distribution~\citep{2018ApJ...856..173T,2023MNRAS.522.5546F}. The effective inspiral spin distribution follows a Gaussian distribution, truncated to the physical range [-1, 1]~\citep{2019MNRAS.484.4216R,2020ApJ...895..128M}. For the merger rate as a function of redshift, we assume the same functional form in $(1+z)$ as the Madau-Dickinson SFR, with a roughly power-law rise at low redshifts, a peak, and roughly power-law decline at high redshifts~\citep{2014ARA&A..52..415M,2020ApJ...896L..32C}.

\begin{figure}
    \centering
    \includegraphics[scale=0.55]{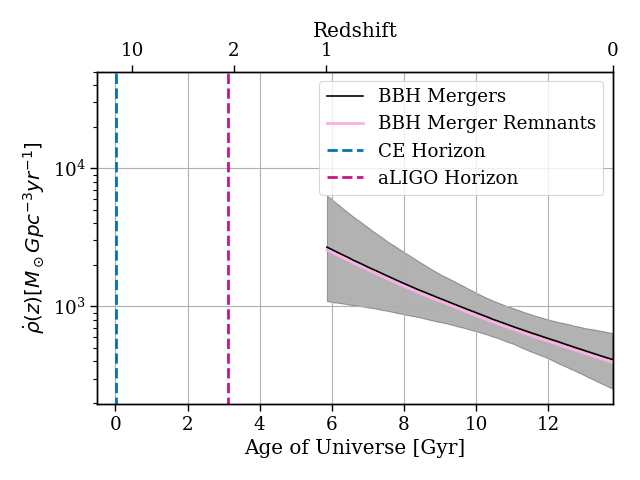}
    \caption{The mass density rate of LVK's stellar mass black holes as a function of cosmic time. The rate increases at larger redshift, or earlier in the universe, but is truncated at $z\sim 1$ due to the depth of LVK observations. The detection horizons for next-generation detectors are shown, with Advanced LIGO at A+ sensitivity extending observation to $z\sim 2$ and Cosmic Explorer (CE) able to measure all mergers in-band out to the beginnings of BBH mergers \citep{2021PhRvD.103l2004H}. The mass density rate of BBH merger remnants is also shown, accounting for the mass that is lost during GW emission. It is indistinguishable from the merger mass density rate within uncertainties. }
    \label{fig:LIGOdensityrate}
\end{figure}

We apply hierarchical Bayesian inference to fit our model to GWTC-3, using the parameter estimation samples~\citep{2021SoftX..1300658A,10.5281_zenodo.5117703,ligo_scientific_collaboration_and_virgo_2021_5546663} and sensitivity estimates~\citep{ligo_scientific_collaboration_and_virgo_2021_5636816} used by~\citet{2023PhRvX..13a1048A}. We compute the mass density rate contained in merging stellar-mass BBHs over cosmic time,  
\begin{equation}\label{eq:rho}
    \dot\rho(z) = \int_{3\,M_\odot}^{100\,M_\odot} \int_{3\,M_\odot}^{m_1} R(z) p(m_1,m_2) (m_1+m_2) dm_2dm_1
\end{equation}
where we have defined the component mass range of stellar-mass BBHs to be $3\,M_\odot < m_2 < m_1 < 100\,M_\odot$.
The computed mass density rate is shown in Fig. \ref{fig:LIGOdensityrate}. Note that this differs slightly from the mass density rate of BBH merger remnants because a fraction of this mass is lost to gravitational radiation. We find that the distinction between the total mass of the component BHs and the mass of the merger remnant is negligible compared to statistical uncertainties, as seen in Fig. \ref{fig:LIGOdensityrate}{, which we calculated using the average between the fits of \cite{2017PhRvD..95b4037H} and \cite{2017PhRvD..95f4024J} as implemented in \texttt{PESummary} \citep{2021SoftX..1500765H}. }

\begin{figure}
    \centering
    \includegraphics[scale=0.55]{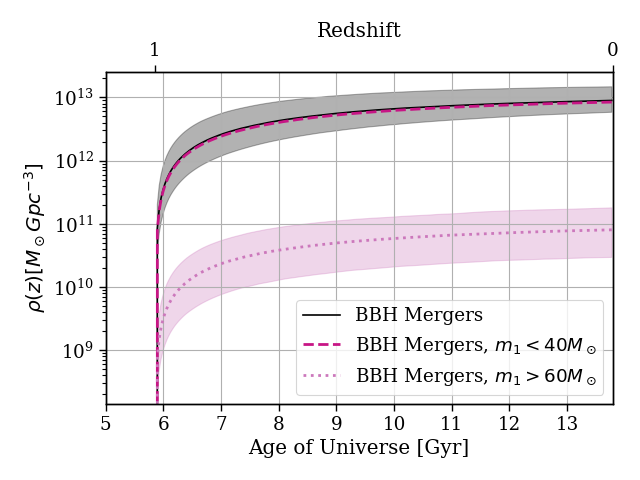}
    \caption{The mass density that has taken part in a merger of LVK's stellar mass black holes over cosmic time. This is computed by taking a cumulative time integral of the mass density rate from Fig.~\ref{fig:LIGOdensityrate}. The cut-off is due to the LVK detection horizon. The mass in mergers where the primary component is $<40 M_\odot$ is indistinguishable from the total mass density on this plot, and the mergers where $m_1>60 M_\odot$ is multiple orders of magnitude smaller than the presented mass density. Thus, the possibility of mass recycling due to hierarchical mergers does not impact our results. }
    \label{fig:LIGOdensity}
\end{figure}

In Fig. \ref{fig:LIGOdensity}, we integrate Eq. \ref{eq:rho} cumulatively over time to produce a mass density with units of $M_\odot \text{Gpc}^{-3}$. This mass density represents the BH mass that has taken part in a BBH merger as a function of redshift (or age of the Universe) between the LVK horizon at $z\sim1$ and today. 
This calculation assumes that every merger observed by LVK adds mass to the `bucket' of merged BHs. However, if hierarchical mergers occur, they would violate this assumption, because hierarchical mergers recycle BH mass. Recent work has found that second-generation mergers and above are only a small subset of all mergers \citep[e.g.][]{2020ApJ...893...35D,2021ApJ...915L..35K,2022ApJ...935L..26F}. 
In order to assess the potential impact of higher-generation mergers, we compute the mass density contained only in BBHs with $m_1 < 40\,M_\odot$. Based on the spin distribution analysis of \cite{2022ApJ...935L..26F}, hierarchical mergers are a negligible contribution to the BBH population with $m_1\lesssim M_\odot$, so we choose a conservative limit $m_1 < 40\,M_\odot$ to ensure that we are capturing exclusively first-generation BBHs. Because mergers with component masses above $40\,M_\odot$ are such a small contribution to the BBH rate, the resulting mass density is indistinguishable within statistical uncertainties from the mass density across all BBH mergers up to $m_1 = 100\,M_\odot$ (see Fig. \ref{fig:LIGOdensity}).
In other words, the mass density of merging BBHs with component masses $>40 M_\odot$ or $>60 M_\odot$ are orders of magnitude smaller than the mass density contained at lower masses. One consequence of this observation is that there is more than enough mass contained in merging BBHs with $m_1 < 40\,M_\odot$ to account for the rate of merging BHs at higher masses, if indeed such higher-mass BHs are hierarchical merger products. We will discuss a similar consideration later in the context of the SMBH mass density.

These data are naturally missing mergers that are beyond the LVK's detection horizon, i.e. those that are farther than $z\sim1$ or occurred before the universe was $\sim 6$ Gyr old. Therefore, Fig.~\ref{fig:LIGOdensity} represents a lower limit of the total mass density in stellar mass BHs that merge over all of cosmic history. To make comparisons to other objects in the universe or make predictions for future detectors, it is necessary to estimate the contribution to the mass density from higher redshifts. Thus, we fit a physical model to the LVK data at lower redshifts that can naturally predict the merger rate at higher redshifts.
\section{How much mass is there in BBH mergers at higher redshifts?}\label{sec:model}

We extrapolate the mass in BBH mergers out to higher redshifts by developing a physically informed model to explain the observed BBH merger rate at $z \lesssim 1$. This allows us to predict the high-redshift merger rate accessible to future gravitational-wave observatories such as Cosmic Explorer and Einstein Telescope~\citep{2020JCAP...03..050M,2021arXiv210909882E}. Our approach also provides a measurement of the metallicity dependence and time delay distribution of BBH mergers, providing insights into their formation histories.

The BBHs that the LVK observes are stellar remnants; consequently, the rate and masses of these BBH mergers depend on the SFR \citep{2024AnP...53600170C}. In this work, we use the UV corrected SFR of \cite{2021ApJ...919...88K}:
\begin{equation} \label{eq:SFR}
    \dot\rho_{\text{SFR}}(t)=0.037t^{1.83}e^{-0.48t}\frac{\text{M}_\odot}{\text{Mpc}^{3} \text{year}}
\end{equation}
which they showed agrees with the \cite{2014ARA&A..52..415M} SFR and matches observations better than their UV+IR SFR. 

Only a fraction of all stellar mass will end up in a merging BBH. We account for this by applying a dimensionless efficiency factor $\epsilon$ to the SFR to give the BBH progenitor formation rate $\dot\rho_{\text{BBHp}}$: 
\begin{equation}\label{eq:e}
    \epsilon = \frac{\dot\rho_{\text{BBHp}}}{\dot\rho_{\text{SFR}}}
\end{equation}
The BBH progenitor formation rate $\dot\rho_{\text{BBHp}}$ accounts for the stellar mass that will eventually contribute to merging BBH systems within some time $\tau_{\text{max}}$, which we take to be the age of the universe.

The efficiency factor $\epsilon$ is not constant with redshift because it likely depends on metallicity \citep{2024AnP...53600170C}. Here we assume that the efficiency as a function of metallicity takes the form of a step function with a constant low metallicity efficiency, $\epsilon_{<Z_{\text{t}}}$, and a constant high metallicity efficiency, $\epsilon_{>Z_{\text{t}}}$ separated by a metallicity threshold $Z_{\text{t}}$.
\begin{equation}\label{e(Z)}
\epsilon(Z) = \left\{
        \begin{array}{ll}
            \epsilon_{<Z_{\text{t}}} & \quad Z \leq Z_{\text{t}} \\
            \epsilon_{>Z_{\text{t}}} & \quad Z \geq Z_{\text{t}}
        \end{array}
    \right.
\end{equation}
We assume that the metallicity threshold is $\log_{10}({Z_{\text{t}}}/{Z_{{\odot}}}) = -0.5$, consistent with many theoretical predictions (see Figure 7 of \citealt{2023ApJ...957L..31F}). While one could fit for the metallicity threshold alongside $\epsilon_{>Z_{\text{t}}}$ and $\epsilon_{<Z_{\text{t}}}$, we find that the current data is unable to provide meaningful constraints, in agreement with previous work \citep{2023arXiv231017625T}.

At each formation redshift (equivalently, formation time), there is a fraction of stars forming with metallicities less than $Z_{\text{t}}$, a fraction of which ($\epsilon_{<Z_{\text{t}}}$) will eventually participate in a BBH merger. Likewise, a fraction of stars have metallicities larger than $Z_{\text{t}}$, a different fraction of which ($\epsilon_{>Z_{\text{t}}}$) will take part in a BBH merger. Therefore, the BBH progenitor formation rate $\dot\rho_{\text{BBHp}}$ depends not only on $Z_{\text{t}}$, $\epsilon_{>Z_{\text{t}}}$, and $\epsilon_{<Z_{\text{t}}}$, but also on the fraction of stars that are below or above $Z_{\text{t}}$:
\begin{equation}\label{progenitor}
    \dot\rho_{\text{BBHp}} = \dot\rho_{\text{SFR}}\left(\epsilon_{<Z_{\text{t}}}\text{f}_{<Z_{\text{t}}} +\epsilon_{>Z_{\text{t}}}\text{f}_{>Z_{\text{t}}}\right)
\end{equation}

We use the metallicity-redshift relation from~\cite{2017ApJ...840...39M} to describe the average metallicity: 
\begin{equation}\label{avgz}
    \langle\log_{10}Z(z)/Z_\odot\rangle = 0.153-0.074z^{1.34}
\end{equation}
We assume that star-forming metallicities at each redshift are drawn from a log-normal distribution centered at Eq.~\ref{avgz} with a scatter of $\sigma=0.4$ dex~\citep{2019MNRAS.490.3740N}.
At each redshift, the low metallicity fraction $\text{f}_{<Z_{\text{t}}}(z)$ is the integral of the log-normal distribution below $\log_{10}({Z_{\text{t}}}/{Z_{{\odot}}}) = -0.5$. 

The BBH merger rate, in turn, must account for the delay time $\tau$ between when a star initially forms to when it eventually takes part in a BBH merger. We assume that the probability distribution of these delay times $p(\tau)$ is a normalized power-law:  
\begin{equation}\label{p(tau)}
    p(\tau) = \frac{\alpha+1}{\tau_{\text{max}} ^{\alpha+1}-\tau_{\text{min}} ^{\alpha+1}} \tau^{\alpha}
\end{equation}
where the power-law index, $\alpha$, defines the steepness of the time delay distribution, with more negative values corresponding to high slopes and therefore a preference for shorter delay times. As mentioned earlier, we fix $\tau_\mathrm{max}$ to the age of the Universe; a different choice is equivalent to a renormalization of the efficiency $\epsilon$. However, $\alpha$ and $\tau_\mathrm{min}$ are free parameters, for which different formation channels and environments produce different predictions. Inferring these delay time distribution parameters from data is therefore an additional way to probe the formation histories of merging BBHs~\citep{2021ApJ...914L..30F}.

Putting everything together, we model the BBH merger rate density as a function of time $t$, where $t$ represents the time elapsed from the beginning of the Universe to merger~\citep{2007PhR...442..166N, 2016PhRvL.116m1102A}:
\begin{equation}\label{eq:bbhmodel}
\begin{split}
    \dot\rho_{\text{BBH}}(t) =& \int_{\tau_{\text{min}}}^{\tau_{\text{max}}} \dot\rho_{\text{BBHp}}(t-\tau) p(\tau)  d\tau.
\end{split}
\end{equation}
We convert between time $t$ and redshift $z$ using Planck 2018 cosmology as implemented in \textsc{Astropy} \citep{2020A&A...641A...6P, 2022ApJ...935..167A}.
Equation~\ref{eq:bbhmodel} depends on the free parameters $\epsilon_{>Z_{\text{t}}}$, $ \epsilon_{<Z_{\text{t}}}$,  $\tau_{\text{min}}$, and $\alpha$, as well as some assumptions about the metallicity-specific SFR (Eqs.~\ref{eq:SFR} and~\ref{avgz}).
We use the method of~\citet{2021ApJ...914L..30F,2023arXiv231203316V} to infer the parameters governing the BBH progenitor formation efficiency and the delay time distribution. We start with the parametric fit to the merger rate as a function of redshift presented in Fig. \ref{fig:LIGOdensityrate}. From this fit, we apply a kernel density estimate (KDE) to the posterior samples for the inferred BBH merger rate at two fixed redshifts, $\dot{\rho}(z = 0)$ and $\dot{\rho}(z = 1)$. We apply a similar KDE to the prior samples. Dividing the posterior KDE by the prior KDE on the merger rate at $z = 0$ and $z = 1$ gives us an approximate likelihood for the efficiency and delay time distribution parameters we wish to fit (listed in Table~\ref{tab:params}). 

We sample from this likelihood using the Python packages \texttt{NumPyro} and \texttt{JAX} \citep{jax2018github, 2018arXiv181009538B,2019arXiv191211554P}. 
The priors on the parameters are all drawn from uniform distributions and are shown in Table \ref{t:table}. We refrain from putting the lower bound on $\alpha$ too low as there are few theoretical predictions for time delay distributions steeper than $\alpha < -1.5$ \citep{2023ApJ...957L..31F}.

\begin{figure*}
    \centering
    \includegraphics[scale=0.5]{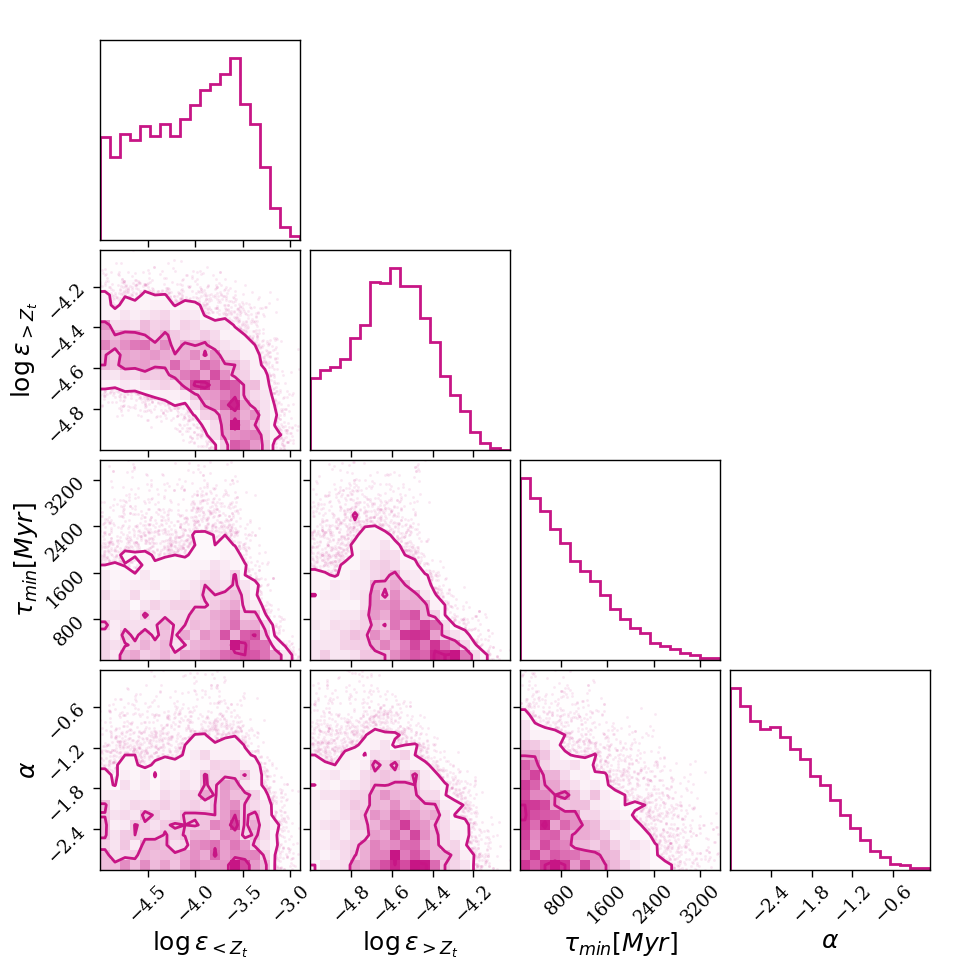}
    \caption{Corner plot from 10,000 posterior samples of the SFR+Efficiency+Time Delay model parameters in Eq. \ref{eq:bbhmodel}. 
    The efficiency at low metallicities is probably higher than at high metallicities (77\% credibility). The time delay parameters rail against the lower prior boundaries, indicating that short delays are favored.}
    \label{fig:corner}
\end{figure*}

Our inferred posteriors on the parameters describing the formation efficiency and delay time distribution are shown in Fig. \ref{fig:corner} and summarized in Table~\ref{tab:params}. The posterior for the high metallicity efficiency, $\epsilon_{>Z_\text{t}}$, peaks at nearly an order of magnitude lower than the low metallicity efficiency, $\epsilon_{<Z_\text{t}}$. We recover that $\epsilon_{<Z_\text{t}} > \epsilon_{>Z_\text{t}}$ at {$77$\% }credibility. 

\begin{table*}
  \caption{\label{tab:params} Progenitor formation efficiency and delay time distribution parameters in our model. We choose uniform priors over all parameters, with minimum and maximum boundaries listed in the Prior column. The Result column reports the posterior $90\%$ credible intervals. The time delay parameters peak at the lower prior edge so the 90th percentile is reported as the upper limit. The results for the efficiencies at high and low metallicities differ by almost an order of magnitude despite having identical priors. The time delay parameters are indicative of short delay times.}
  \centering 
  \begin{threeparttable}\label{t:table}
    \begin{tabular}{cccc}
    Parameter  & Description  & Prior & {Result}\\
     \midrule\midrule
    $\log_{10}\epsilon_{<Z_{\text{t}}}$  &   Low Metallicity Efficiency      &   U(-5,-1) & $-3.99^{+0.68}_{-0.87}$                
    \\
    \cmidrule(l  r ){1-4}
    $ \log_{10}\epsilon_{>Z_{\text{t}}}$& High Metallicity Efficiency  & U(-5,-1) & $-4.60^{+0.30}_{-0.34}$ 
    \\ 
    \cmidrule(l r ){1-4}
     $\tau_{\text{min}}$ & Minimum Time Delay & U(100,5000) Myr & $<1.9$ Gyr  \\ 
    \cmidrule(l r ){1-4}
    $\alpha$ &  Time Delay Distribution Power-Law Index  & U(-3,0) & $<-1.32$   \\
    \midrule\midrule
    \end{tabular}
\end{threeparttable}
  \end{table*}

We also find a preference for short delay times, with a minimum delay time $\tau_\mathrm{min}<1.9$ Gyr and power-law index $\alpha<-1.32$ at 90\% credibility. The posteriors for both of these parameters peak at even lower values, {with $\tau_\mathrm{min}$ and $\alpha$ railing against their lower prior boundary at $\tau_\mathrm{min} = 0.1$ Gyr and $\alpha = -3$.} 
The joint posterior on $\alpha$ and $\tau_\mathrm{min}$ follows the expected correlation, so that larger (less negative) values of $\alpha$ are more strongly ruled out for larger values of $\tau_\mathrm{min}$ and vice versa.
These results are consistent with previous findings that short delay times are preferred, driven by the fact that the shape of the BBH merger rate resembles the SFR at $z < 1$~\citep{2021ApJ...914L..30F,2023arXiv231017625T,2023ApJ...957L..31F,2023arXiv231203316V}. 


\begin{figure}
    \centering
    \includegraphics[scale=0.55]{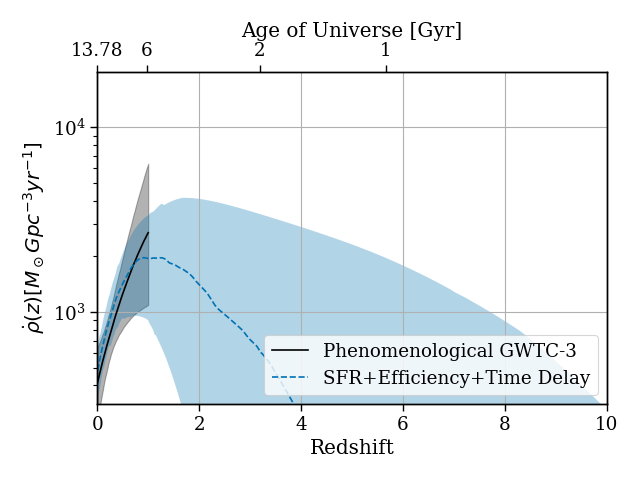}
    \caption{The rate density of stellar mass BBH mergers as a function of redshift as inferred from GWTC-3. The black line shows the phenomenological model, and is the same as in Fig. \ref{fig:LIGOdensityrate}. The blue shows the SFR+Efficiency+Time delay model of Eq.~\ref{eq:bbhmodel}, which is able to extrapolate to higher redshifts beyond the LVK detection horizon. Shaded bands denote 90\% credible regions.}
    \label{fig:ratemodel}
\end{figure}

Figure~\ref{fig:ratemodel} shows our inference for the BBH merger rate with the SFR+Efficiency+Time Delay model (blue), compared to the phenomenological model of Fig.~\ref{fig:LIGOdensityrate}. 
The expected turnover in the merger rate due to the peak star formation is seen in the SFR+Efficiency+Time Delay model, which is extrapolated out to high redshifts, but is not yet observed for current detections due to the detector horizon. Our results in this section depend on the SFR as a function of redshift and metallicity. The SFR is highly uncertain, especially at low metallicities and high redshifts beyond the peak of star formation. Indeed, future GW observations of high-redshift BBH mergers are promising tools to constrain the SFR itself~\citep{2019ApJ...886L...1V}. Nevertheless, we do not expect the current SFR uncertainties to qualitatively change our conclusions that low metallicity formation environments, coupled with short delay times, are preferred. \citet{2021ApJ...914L..30F} considered different prescriptions for the metallicity distribution as a function of redshift and found that the delay time inference was fairly robust to these different choices. As the GW statistical uncertainties shrink, it will become critical to marginalize over uncertainties in the metallicity-specific SFR in future work.

By modelling the BBH merger mass density rate, we have fit parameters corresponding to their formation. Now, we use this model to compare to other stellar and black hole objects in the universe.
\section{Comparison to stellar progenitors and supermassive black holes}
\label{sec:comparison}

In this section, we compare the inferred mass density in stellar-mass BBHs to the mass density of high-mass stars as well as the mass density in supermassive black holes (SMBHs). High-mass stars are thought to be the progenitors of stellar-mass BBHs. Meanwhile, stellar-mass BBHs may contribute to seeding the earliest, highest-redshift SMBHs. Comparing the mass contained in these various objects across cosmic time is a first step towards understanding their possible connection. \maya{Our results in this section can be compared to \citet{2022ApJ...924...56S}, particularly their Fig. 7, which shows their computed mass densities of BHs across redshift. Our orange "All Stellar Mass $>10M_\odot$" line in Fig. \ref{fig:densitymodel} agrees within an order of magnitude to their black hole mass density which includes all black holes. While they perform stellar population synthesis simulations to predict the mass contained in different populations of stellar-mass BHs, our results are derived directly from measurements of the SFR and the BBH merger rate.}

\subsection{Mass Density of High Mass Stars}
The massive stars that precede BHs are rare and short-lived, but BHs exist in perpetuity and can be a useful probe of their predecessors. We compare the mass density in merging BBHs to the mass density in high-mass ($>$10 $M_\odot$) stars, since only high-mass stars evolve into BHs \citep{2023arXiv230409350H}.

To estimate the formation histories of high-mass stars, we must make some assumption about the initial mass function (IMF). We adopt the IMF of \cite{2021sfrg.book.....Z}: 
\begin{equation}\label{imf}
    \xi(m) \propto \left(\frac{m}{0.5}\right)^{-2.3\pm0.36}
\end{equation}
Eq. \ref{imf} is valid between 0.5 $M_\odot$ and 150 $M_\odot$, so we only consider stars in this mass range. We assume a constant IMF across all redshifts. Although this assumption is likely to break down due to, for example, the metallicity dependence of the IMF~\citep{2013pss5.book..115K,2018PASA...35...39H}, the high-mass SFR is better measured than the IMF or SFR independently~\citep{2018A&A...619A..77K,2020A&A...636A..10C}.

\begin{figure*}
    \centering
    \includegraphics[scale=0.6]{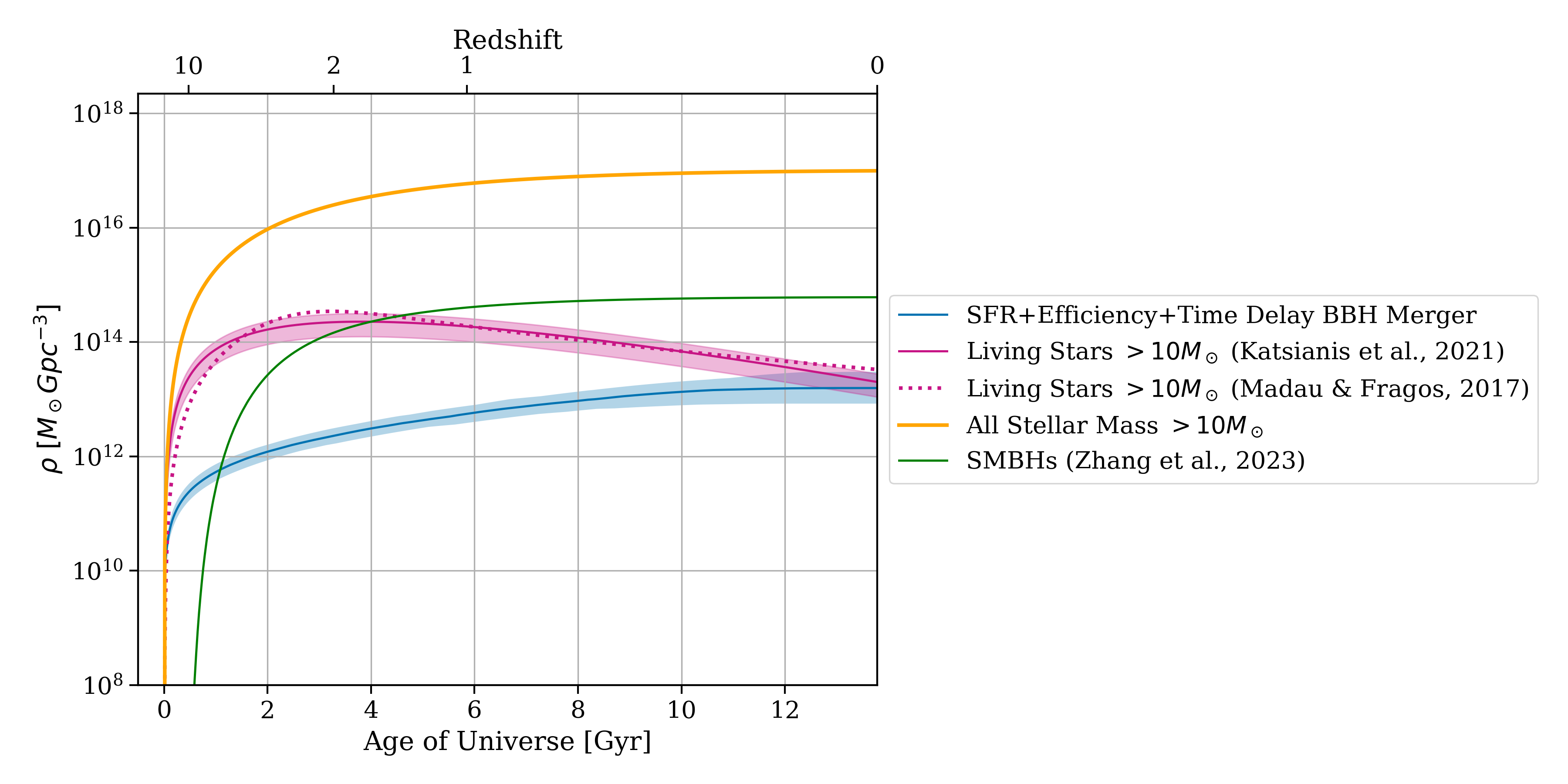}
    \caption{Mass density as a function of redshift/age of the universe. 
    The cumulative integral of our model for SFR+Efficiency+Time delay from Fig. \ref{fig:ratemodel} is shown in blue. The mass density of $>10M_\odot$ stars that are currently alive at any redshift is shown in pink with the uncertainty band due to the IMF in Eq. \ref{imf} and the pink dotted line showing an alternative SFR. The overlap in the regions of living massive stars and BBHs indicate that there may be more mass in merging BHs than there is in their living progenitors. In orange the cumulative mass stellar mass density using the \cite{2021ApJ...919...88K} SFR, and accounting for mass lost with a return fraction of 0.27 as specified in \citep{2014ARA&A..52..415M} is shown. The green line shows the cumulative mass density of SMBHs from the \textsc{trinity} simulation \citep{2023MNRAS.518.2123Z}. The credible region is not shown as it is too small to see. In the early universe there was enough stellar mass BHs merging to account for the mass budget of SMBHs. }
    \label{fig:densitymodel}
\end{figure*}

{The total stellar mass density for stars $>10 M_\odot$, including stellar remnants, is given by the cumulative integral of the SFR, for which we adopt the \cite{2021ApJ...919...88K} SFR (Eq.~\ref{eq:SFR}), multiplied by the fraction of high mass stars. We also subtract the mass lost during stellar life, given by the return fraction $R$. We assume $R = 0.27$~\citep{2014ARA&A..52..415M}. The resulting mass density of high-mass stars is shown by the orange line in Fig.~\ref{fig:densitymodel}.}

Next we calculate the mass density in high-mass stars that are alive at a particular time. To do so, we consider the stellar lifetime as a function of mass~\citep{1988asco.book.....H}:
%
\begin{equation}
    t_{\text{lifetime}} = 10^{10}(M/M_\odot)^{-2.5}\,\text{years}.
\end{equation}
Computationally, we divide stellar masses into bins of size 10 $M_\odot$ and consider the average lifetime in each mass bin. At each time $t$, we evaluate the mass density of living stars in each mass bin:
\begin{equation}
    \rho = \int_{t-t_{\text{lifetime}}}^{t} \dot\rho_{\text{SFR}}(t') dt'
\end{equation}
%
We then sum over all mass bins to give the mass density in all living massive stars as a function of redshift/age of the universe. 
The mass density in living high-mass stars, assuming the \cite{2021ApJ...919...88K} SFR given in Eq. \ref{eq:SFR}, is shown in the pink solid line of Fig.~\ref{fig:densitymodel}.
The uncertainty band represents the uncertainty in the IMF.
We also repeat the calculation with the SFR of \cite{2017ApJ...840...39M}:
%
\begin{equation}\label{eq:madauSFR}
    \dot\rho_\mathrm{SFR}(z) =  0.01\frac{(1+z)^{2.6}}{1+[\frac{1+z}{3.2}]^{6.2}} \frac{\text{M}_\odot}{\text{Mpc}^{3} \text{year}}
\end{equation}
to gauge the possible uncertainties due to choice of SFR. 

Fig.~\ref{fig:densitymodel} compares the total stellar mass density of high-mass stars (orange), as well as the contribution from living stars (pink), against the mass density in merging BBHs (blue). The mass density in merging BBHs is the cumulative time integral of our inferred BBH merger rate shown in Fig. \ref{fig:ratemodel}.
Merging BBHs are extremely rare end states of massive stellar evolution, making up only $\sim0.01\%$ of the high-mass stellar density when we account for stellar remnants. However, massive stars are short lived, so that as early as $\sim 2.5$ Gyr ago, there may have been more mass in merging BBHs as there is in living high-mass stars. This highlights that merging BBHs are a vital tool for studying massive stars.

Currently, our inferred BBH merger mass density at high redshifts is an informed extrapolation of the low-redshift data. By construction, our model assumes that star formation preceded BBH mergers. Future GW detectors will be able to directly measure the mass in BBH mergers at high redshifts, while high-redshift galaxy observations will better constrain the SFR at these early times. If the high-redshift mass density of BBHs is found to be larger than the {cumulative stellar mass} at these early times, it would be a compelling argument for primordial BHs.

\subsection{Mass Density in Supermassive Black Holes}\label{sec:SMBH}
To what degree could stellar mass BHs contribute to the growth of SMBHs? The population of SMBHs with masses as large as $10^9 M_\odot$ reside in the centres of most galaxies and there is evidence for them as early as $z=7$ \citep{2020ARA&A..58...27I}. There are multiple proposed methods of forming and growing these behemoths. In principle, BHs of masses $10-100 M_\odot$ could be low mass seeds for SMBHs if they subsequently undergo super-Eddington accretion and/or have runaway mergers with other compact objects \citep{2016MNRAS.462.3812T, 2020ARA&A..58...27I,2022ApJ...924...56S}. One source of growth may be hierarchical mergers of stellar mass BBHs in dense environments \citep{2020ARA&A..58...27I}. 

In Fig. \ref{fig:densitymodel} we plot the cosmic SMBH mass density from \textsc{trinity}, a model relating haloes, galaxies, and their SMBHs, alongside the previously discussed mass densities \citep{2023MNRAS.518.2123Z}. According to our SFR+Efficiency+Time Delay extrapolation of the local BBH merger rate, in the first $\sim1$ Gyr of the Universe, the mass in stellar mass BHs was orders of magnitude larger than that in the young SMBHs. This is consistent with stellar mass BHs serving as seeds for SMBHs, but only at very early times. {Our conclusions depend on our assumptions about the SFR, which suffers from observational uncertainties. We note that our SFR model from~\citet{2021ApJ...919...88K} assumes continuous star formation starting at the beginning of the Universe, rather than adopting a maximum redshift. However, our results do not noticeably change if we instead assume that there is no star formation prior to $z=10$.}

\begin{figure}
    \centering
    \includegraphics[scale=0.55]{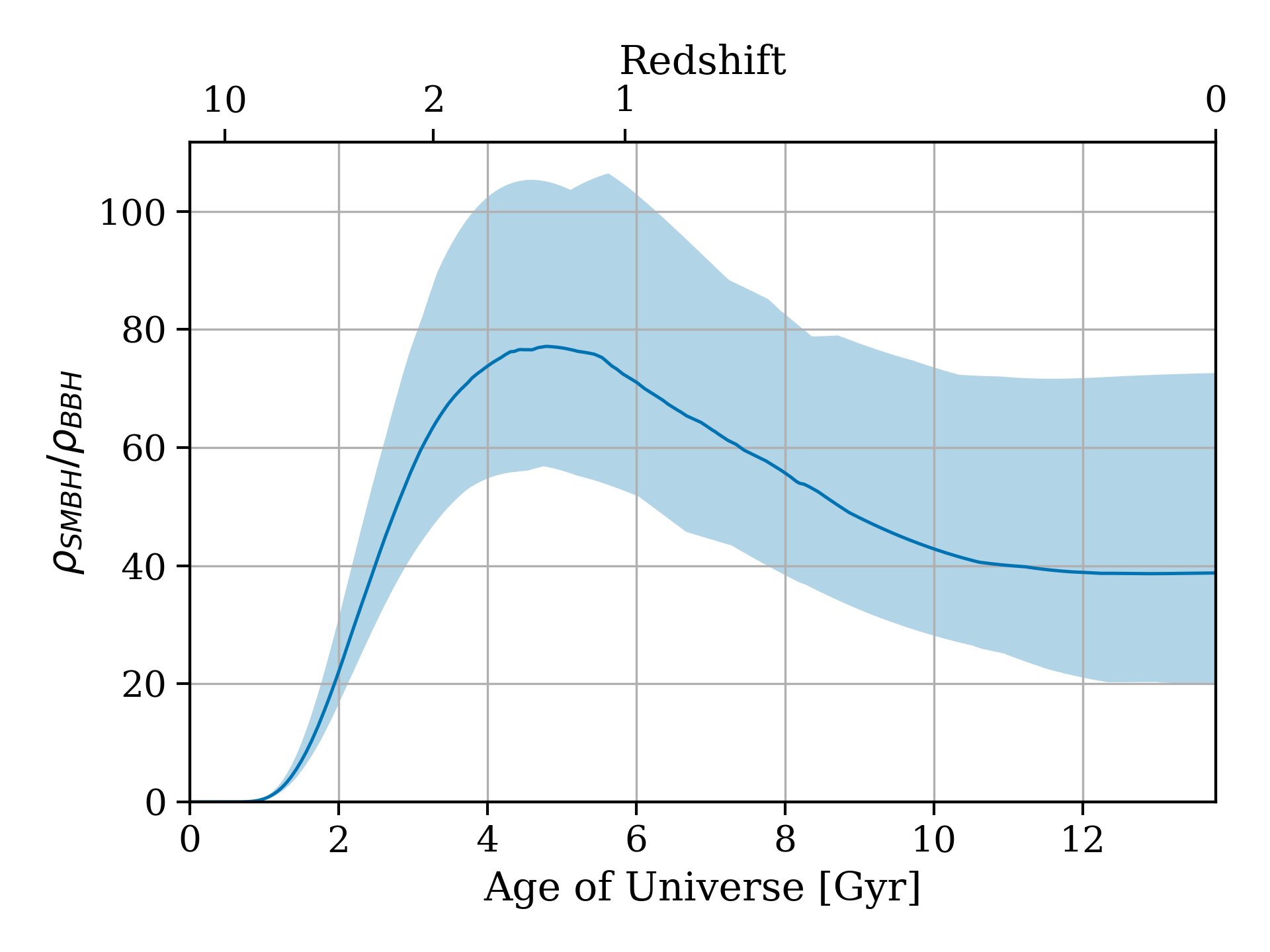}
    \caption{The ratio of SMBH mass density compared to the merging stellar mass BBH mass density. Prior to 1 Gyr, there was more mass in merging BBHs according to our model, but shortly after this time the mass in SMBHs was as much as $\sim 110$ times larger. Today the SMBH mass density remains $\sim 40$ times larger than that in BBH mergers. }
    \label{fig:SMBHeff}
\end{figure}

After the Universe is $\sim 1$ Gyr old, SMBHs begin to grow rapidly. Merging stellar mass BBHs alone can no longer explain the mass budget of SMBHs, unless the BBH merger efficiency $\epsilon$ was much higher than our model assumes, either due to a higher merger frequency or a preference for higher mass BBHs compared to the local BBH mass distribution. {The efficiency $\epsilon$ would need to be $\sim 50-110$ times higher in the early universe to account for all the mass in SMBHs (see Fig. \ref{fig:SMBHeff}). Even this efficiency estimate requires that all BBH mass goes into SMBHs, which is extremely improbable given that we know hierarchical mergers are rare today.} Nonetheless, if future GW detectors find that BBH mergers are more common or more massive in the early universe, then there could be a compelling connection to SMBHs. Runaway BBH mergers would result in merging intermediate mass black holes (IMBHs), which may also be observable by future GW detectors that are sensitive at lower frequencies than the LVK. Therefore, studying the mass densities of different BH populations across redshift could be a useful way to connect multiband GW detections, from nanohertz to kilohertz frequencies~\citep{2022arXiv220204764C,2023MNRAS.525.2851S,2022ApJ...924...56S}. 
Our framework could be a useful pathway to not only learning about the formation of LVK's BBHs, but perhaps SMBHs as well.

\section{Conclusion}

There is a plethora of open problems that remain to be answered by GWs, including the connection between stars, merging BBHs, and possibly SMBHs. 
We have computed the mass density rate of LVK's BBHs and extrapolated it to higher redshifts by inferring the efficiency by which BBHs are made from stars and the delay time from their formation to their merger. 
We found the efficiency to be $10^{-4.60^{+0.30}_{-0.34}}$ in high metallicity environments and $10^{-3.99^{+0.68}_{-0.87}}$ in low metallicity environments (90\% credibility). Previous work found that individual GW events likely came from low metallicity environments within a population synthesis framework \citep{2021ApJ...914L..32A,2023ApJ...950..181W}. This work is the first time that a low metallicity preference has been shown directly from the inferred merger rate as a function of redshift as opposed to targeted analysis of individual events. For a power-law delay time distribution, we inferred a minimum time delay that is $<1.9$ Gyr and a power law slope that is $<-1.32$ (90\% upper limits). Hence, the BBH merger rate is consistent with BBHs that form preferentially at low metallicities, while  preferring short time delays. This does not necessarily mean the observed BBHs come from metal-poor regions nearby, since many BBH progenitors may have formed in the distant universe with a long time delay drawn from the tail of our distribution~\citep{2023ApJ...957L..31F}. Our results suggest that the delay time distributions may be steeper than typical predictions for BBH formation mechanisms.
For example, \citet{2023ApJ...957L..31F} report typical predictions of $\alpha = -1$ for isolated binary evolution with a common envelope phase, and $\alpha = -0.35$ for isolated binary evolution with only stable mass transfer. For dynamically assembled BBH mergers in globular clusters, \citet{2024arXiv240212444Y} report delay time distributions with $-1.2 < \alpha < -0.8$ depending on the mass of the BBH merger. However, the tension with these predictions is not yet significant. In particular, Fig. \ref{fig:corner} shows that the shortest $\tau_{\text{min}}$ are still consistent with $\alpha=-1$.

Using this SFR+Efficiency+Time Delay model, we extrapolated the BBH merger rate to high redshifts and proposed using the cumulative mass densities of merging BBHs to find connections to other stellar and BH populations. We found that at an age of the universe as early as $\sim$ 11 Gyr, there may be more mass in merging stellar mass BBHs than in massive stars $>10\,M_\odot$ that are still alive. However, merging BBHs are only a small fraction of all BHs. If we include the mass in massive stellar remnants, merging BBHs make up $\sim0.01\%$ of all stellar mass $>10\,M_\odot$ today. 
By assumption, there should always be more total stellar mass than mass in BBHs that are merging. Once future GW detectors are able to probe very early times, this provides a natural test for stellar versus primordial BH origins, similar to the cosmological tests performed with high-redshift galaxy and quasar observations~\citep{2018MNRAS.477.5382B,2023NatAs...7..731B}. 

Likewise, we compared the mass in merging stellar-mass BBHs to the mass density in SMBHs. According to our model, we found that early in the universe, before $\sim 1$ Gyr, there was more than enough mass in merging BBHs to account for the mass in young SMBHs. These first BBHs may have contributed to the humble beginnings of the gargantuans observed today. However, they are unlikely to serve as the dominant mass source, particularly after the first Gyr. Nevertheless, future GW observations will be able to directly probe the populations of stellar-mass BBH mergers that may serve, or be closely related to, the seeds of SMBHs. 

The gravitational wave catalog currently available to us is already providing information on the high redshift universe. This information will only get better as more events are observed and detector sensitivity increases. 

\section*{Acknowledgements}

We thank Tom Callister, Reed Essick, Daniel Holz and Aditya Vijaykumar for helpful conversations, and Matthew Mould for his comments on the manuscript. AS-Z acknowledges support from the Natural Sciences and Engineering Research Council of Canada (NSERC) and the Ontario Graduate Scholarship. AS-Z and MF are supported by NSERC RGPIN-2023-05511. This material is based upon work supported by NSF's LIGO Laboratory which is a major facility fully funded by the National Science Foundation.

This work made use of the open source Python packages \texttt{Astropy} \citep{2022ApJ...935..167A}, \texttt{matplotlib} \citep{2007CSE.....9...90H}, and \texttt{corner} \citep{2016JOSS....1...24F}.

\bibliography{newlibrary}{}
\bibliographystyle{aasjournal}

\end{document}